\documentclass[journal]{IEEEtran}
\usepackage{epsfig,graphics,epsf,color,amsmath,balance,cite,authblk}
\usepackage{graphicx}
\usepackage{psfrag}
\usepackage{stfloats}
\usepackage{amsmath}
\usepackage{amssymb}
\usepackage{array}

\begin{document}

\title{Joint Shortening and Puncturing Optimization for Structured LDPC Codes}
\author{
\authorblockN{Yuejun~Wei, Yuhang~Yang, Ming~Jiang, Wen~Chen and Lili~Wei}
\thanks{Y. Wei, Y. Yang, W. Chen and L. Wei are with the Department of Electronic Engineering,
Shanghai Jiao Tong University, Shanghai, China (e-mail: \{yjwei;yhyang;wenchen;liliwei\}@sjtu.edu.cn); M. Jiang is with National Mobile Communications Research Laboratory, Southeast University, Nanjing, China (e-mail: jiang\_ming@seu.edu.cn).}
\thanks{This paper is supported by the National 973 Project \#2012CB316106, and \#2009CB824904,
and by NSF China \#60972031 and \#61161130529.}
}\maketitle
%%%%%%%%%%%%%%%%%%%%%%%%%%%%%%%%%%%%%%%%%%%%%%%%%%%%%%%%%%%%%%%%%%%%%%%%%%%%%%%%%%%%%%%%%%%%%%%%%%%%%%%%%%%%%%%%%%%%%%%%%%%%%%%%%%%%%%%%%%%%%%%%%%%%%%%%%%%%%%%%
\begin{abstract}
The demand for flexible broadband wireless services makes the pruning technique, including both shortening and puncturing, an indispensable component of error correcting codes. The analysis of the pruning process for structured low-density parity-check (LDPC) codes can be considerably simplified with their equivalent representations through base-matrices or protographs. In this letter, we evaluate the thresholds of the pruned base-matrices by using protograph based on extrinsic information transfer (PEXIT). We also provide an efficient method to optimize the pruning patterns, which can significantly improve the thresholds of both the full-length patterns and the sub-patterns. Numerical results show that the structured LDPC codes pruned by the improved patterns outperform those with the existing patterns.
\end{abstract}
%%%%%%%%%%%%%%%%%%%%%%%%%%%%%%%%%%%%%%%%%%%%%%%%%%%%%%%%%%%%%%%%%%%%%%%%%%%%%%%%%%%%%%%%%%%%%%%%%%%%%%%%%%%%%%%%%%%%%%%%%%%%%%%%%%%%%%%%%%%%%%%%%%%%%%%%%%%%%%%%
%%%%%%%%%%%%%%%%%%%%%%%%%%%%%%%%%%%%%%%%%%%%%%%%%%%%%%%%%%%%%%%%%%%%%%%%%%%%%%%%%%%%%%%%%%%%%%%%%%%%%%%%%%%%%%%%%%%%%%%%%%%%%%%%%%%%%%%%%%%%%%%%%%%%%%%%%%%%%%%%
\section{Introduction}
\PARstart{T}{he} structured low-density parity-check (LDPC) codes have been widely used in current communications standards, such as IEEE 802.16e \cite{IEEE_P802_16e} and IEEE 802.11n \cite{IEEE_P802_11n} standards. In these standards, one base-matrix of LDPC codes is designed specifically for one code length and rate in order to achieve better performance. The number of the transmitted bits is determined by many issues, such as bandwidth and modulation, and can be an arbitrary value around the coding lengths defined in the standards. Therefore, pruning techniques are utilized to make the transmission rates and lengths more flexible.

The pruning of the LDPC codes usually consists of two operations, shortening and puncturing, which have already been extensively studied for the binary \cite{Tian2005EURASIP,Richter2006ISIT,Ha2004IT} and the non-binary \cite{Klinc2008Allerton} LDPC codes. Many design schemes for the rate-compatible LDPC codes \cite{Pishro-Nik200702IT,Klinc2005Globecom} are also proposed using efficient shortening and puncturing techniques, where the degree distributions of the irregular LDPC codes, the shortened bits and the punctured bits can be optimized through density evolution and differential evolution \cite{Richardson2001IT}, respectively.

Systematic methods for selecting puncturing patterns include classification of k-step recoverable nodes \cite{Ha200602IT}, puncturing degree 2 nodes \cite{Kim200902TCOM} and progressive node puncturing while minimizing the average number of punctured nodes connected to a check \cite{El-Khamy2009JSAC}. In \cite{Nguyen2010Allerton}, higher rate codes are not obtained by puncturing but rather by extending the information part of the parity check matrix. Besides the stopping sets, there remains some other criteria for the selection of the punctured bits, usually focusing on the distributions or the distances between the punctured bits in Tanner graph \cite{Vellambi2006ISIT,Vellambi200902TCOM,Wu2009Globecom}. Similarly, the shortening pattern of the coded bits are also developed in \cite{Milenkovic200608IT} and \cite{Liu2009CL}.

Further extensions of the density evolution, such as the multi-edge type density evolution \cite{RiU04b} and the protograph based extrinsic information transfer (PEXIT) \cite{Liva2007Globecom}, have been shown to perform more efficiently. Especially, the shortening \cite{Liu2009CL} and the puncturing \cite{Wu2009Globecom} of structured LDPC codes can be easily analyzed due to the simple representations of the base-matrices. We utilize the PEXIT to analyze the pruning of structured LDPC codes, where the shortening can be easily performed by column erasure and the puncturing is inherent in the PEXIT analysis.
%The connections between the variable nodes and check nodes are considered in PEXIT
%analysis. Hence, the effects of the trapping sets caused by puncturing can also be
%evaluated for the protographs, which is mapped from the base-matrices without multi-edges.

In this letter, we aim to jointly optimize the shortening and puncturing patterns for the structured LDPC codes, which can generate more good codes with different rates and lengths from the finite codes defined in IEEE 802.11n standard. The non-greedy search algorithm is employed to optimize the joint shortening and puncturing patterns. We propose a T-stage optimization approach that progressively selects the shortened nodes and the punctured nodes according to the thresholds calculated by PEXIT, which can effectively avoid the performance loss due to unilaterally selecting shortening or puncturing columns. Numeric analysis and simulation results show that the improved patterns obtained from the proposed method can achieve noticeable performance gain over both the pruning schemes in IEEE 802.11n and the combining schemes with the shortening and puncturing patterns described in \cite{Wu2009Globecom} and \cite{Liu2009CL}.

%In this letter, we aim to jointly optimize the shortening and puncturing patterns using an efficient ranking criterion proposed in \cite{Wu2009Globecom}, where the shortened nodes and punctured nodes are progressively selected step by step according to the thresholds calculated by PEXIT. Numeric thresholds and simulation results show that the optimized patterns obtained from the proposed method can achieve noticeable performance gain over both the pruning schemes in IEEE 802.11n and the combining schemes with the shortening and puncturing patterns described in \cite{Wu2009Globecom} and \cite{Liu2009CL}.
%
%The rest of the paper is as follows. Section II introduces the pruning schemes for structured LDPC codes according to the base-matrices. In Section III, we formulate the progressive optimization for searching the optimized pruning patterns. The comparisons of thresholds and error performance between different pruning patterns are presented in section IV. Finally, a brief summary is given in section V.

\section{Pruning for Structured LDPC Codes}
The pruning technique is usually combined with the shortening and puncturing. The shortening is achieved by placing some information bits known for both the transmitter and receiver. The bits to be shortened usually can be set to all ones or all zeros before encoding, while the reliabilties on the corresponding bits are set to infinity in the decoder. On the other hand, puncturing is to keep some bits in the coded sequence not to be transmitted and punctured bits are regarded as erased symbols in the decoder. The shortened information bits and punctured coded bits together determine the final transmission rate and length.

\begin{figure*}[!t]
\normalsize
\centerline{\psfig{file=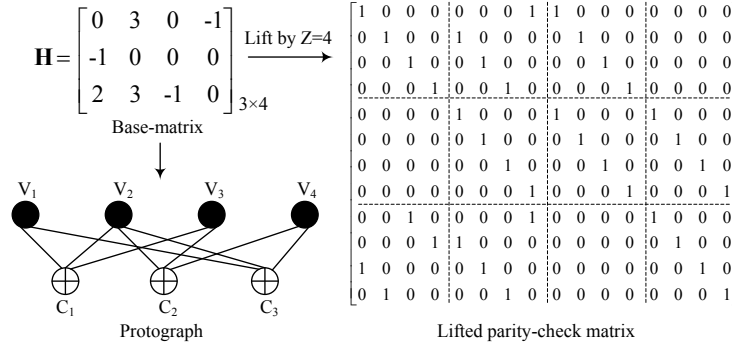,width=3.9in}}
\centering
\caption{The examples of lifting and mapping from the base-matrix. \label{fig:LDPC-BM-Protograph}}
\hrulefill
\end{figure*}

%\begin{figure}[bpht]
%    \begin{center}
%      \epsfxsize=7.0in\includegraphics[scale=0.62]{LDPC-BM-Protograph.eps}
%      \caption{The examples of lifting and mapping from the base-matrix\label{fig:LDPC-BM-Protograph}}
%    \end{center}
%\end{figure}

The descriptions of the pruning schemes for structured LDPC codes can be simplified by using base-matrices with lifting factor $Z$ \cite{IEEE_P802_16e,IEEE_P802_11n,Sridhara2001CISS}. Fig. \ref{fig:LDPC-BM-Protograph} shows an example of the mapping between a structured LDPC code and its base-matrix. In this example, a rate 1/4 base-matrix ${\bf{H}}$ is given, where the lifting factor $Z$ is set to 4 and the columns and rows of the base-matrix are set to $n=4$ and $m=3$, respectively. The entries $h_{i,j}$ of the base-matrix are mapped to all-zero matrices or circulant permutation matrices of size $4 \times 4$, when $h_{i,j}$ is set to -1 or 0, 1, 2, 3, respectively. The circulant permutation matrix, mapped from the entry $h_{i,j}\ge 0$, is right shifted from the identity matrix $\bf{I}$ by $h_{i,j}$ times. The lengths of the lifted coded bits and information bits are $N=n\times Z=16$ and $K=k\times Z=4, k=n-m=1$, respectively.

It is straightforward to define the pruning patterns in terms of the base-matrix by the indices of the shortened/punctured column vectors. We denote the detailed shortening and puncturing patterns as the index ensembles $S_\alpha=\{s_1, s_2, ..., s_\alpha\}$ and $P_\beta=\{p_1, p_2, ..., p_\beta\}$, respectively. The $j$-th column vector of the base-matrix is to be shortened or punctured, if the index $j$ belongs to $S_\alpha$ or $P_\beta$. Then, the original base-matrix ${\bf{H}}$ is transformed into ${\bf{H}}_{\alpha; \beta}$ by the pruning pattern
\begin{equation}\label{pruning-pattern}
{\bf{\Pi}}\{\alpha; \beta\} = S_\alpha \cup P_\beta = \{s_1, s_2, ..., s_\alpha; p_1, p_2, ..., p_\beta\},
\end{equation}
where the parameters $n$ and $k$ are decreased to $n_{\alpha; \beta} = n - \alpha - \beta$ and $k_{\alpha} = k - \alpha$, respectively. The transmission rate of the pruned matrix ${\bf{H}}_{\alpha; \beta}$ turns into $(k-\alpha)/(n-\alpha-\beta)$.

The pattern ${\bf{\Pi}}\{\alpha; \beta\}$ not only represents the column vectors that can be shortened and punctured, but also determines the priorities of the column vectors to be selected from the patterns. Assuming that $N_s$ and $N_p$ bits of the entire coded block are shortened and punctured, respectively, then totally $\alpha + \beta$ columns will be selected to be shortened and punctured, where $\alpha$ and $\beta$ are equal to $\left\lceil {{N_s}/{Z}} \right\rceil$ and $\left\lceil {{N_p}/{Z}} \right\rceil$, respectively. All coded bits corresponding to the first $\alpha - 1$ and $\beta - 1$ columns in the ensembles $S_\alpha$ and $P_\beta$ are shortened and punctured, respectively. Then, the remaining $N_s - (\alpha - 1) \times Z$ and $N_p - (\beta - 1) \times Z$ bits will be sequentially selected from the $s_\alpha$-th and $p_\beta$-th columns in the ensembles $S_\alpha$ and $P_\beta$, respectively. After such pruning processes for the base-matrix ${\bf{H}}$, the total length of the transmitted bits is $N = n \times Z - N_s - N_p$ and the practical transmission rate is changed into $(k \times Z - N_s)/N$.
%%%%%%%%%%%%%%%%%%%%%%%%%%%%%%%%%%%%%%%%%%%%%%%%%%%%%%%%%%%%%%%%%%%%%%%%%%%%%%%%%%%%%%%%%%%%%%%%%%%%%%%%%%%%%%%%%%%%%%%%%%%%%%%%%%%%%%%%%%%%%%%%%%%%%%%%%%%%%%%%
%%%%%%%%%%%%%%%%%%%%%%%%%%%%%%%%%%%%%%%%%%%%%%%%%%%%%%%%%%%%%%%%%%%%%%%%%%%%%%%%%%%%%%%%%%%%%%%%%%%%%%%%%%%%%%%%%%%%%%%%%%%%%%%%%%%%%%%%%%%%%%%%%%%%%%%%%%%%%%%%
\section{Optimization Using Non-Greedy Ranking}
Any pruning pattern can be evaluated via PEXIT analysis \cite{Liva2007Globecom}, given a structured LDPC code with a certain base-matrix. However, it is almost impossible for a pruning pattern to guarantee that there is always a sub-pattern with the minimum threshold for the pruning length. Thus, finding globally optimal pruning pattern of a base-matrix is very difficult for the exhaustive search. An efficient search algorithm for structured LDPC codes is introduced in \cite{Wu2009Globecom}, where a non-greedy (NG) ranking criterion is utilized.

Here, we progressively search the shortened and punctured columns step by step according to the NG ranking criterion, in order to avoid the performance loss due to unilaterally selecting shortening or puncturing columns at first. Assume that the maximum length of the shortened/punctured columns is $\alpha = \beta = T$. Hence, the maximum length of the shortened and punctured bits in total is $N_p + N_s = 2T \times Z$. Then, we can transform the optimization problem into a $T$-stage process by NG ranking criterion. In the $t$-stage process, the $i$-th pruning pattern with length $\alpha=\beta=1$ is defined as follows:
\begin{equation}\label{patterns-t-stage}
{\bf{\Pi}}^{t,i}\{1;1\} = S^{t,i}_1 \cup P^{t,i}_1 = \{s^{t,i}_1; p^{t,i}_1\}, 1 \le t \le T.
\end{equation}
The proposed optimization procedure starts from the exhaustive search of the temporary optimal pattern for the base-matrix ${\bf{H}}$ in the $1$-stage. The number of candidate columns is $k$, when we determine the index of the shortened column at first. The punctured column is selected from the remaining $n-1$ column vectors in the shortened base-matrix. So we define the set of candidate pruning patterns in the $1$-stage optimization as ${\bf{\Omega}}_1={\bf{\Pi}}^{1,i}\{1;1\}, i=1, 2, ..., |{\bf{\Omega}}_1|$, where $|{\bf{\Omega}}_1|$ is equal to $k \times (n-1)$.

The threshold $\gamma^{1,i}_{1;1}$ of the base-matrix ${\bf{H}}^{1,i}_{1;1}$ pruned by the pattern ${\bf{\Pi}}^{1,i}\{1;1\}$ can be evaluated by PEXIT analysis. The pruning patterns with the lowest $\tau$ thresholds in the set ${\bf{\Omega}}_1$ are reserved and form subset for the next stage optimization, denoted by $\hat {\bf{\Omega}}_1=\{{\bf{\Pi}}^{1,i}\{1;1\}, i=1, 2, ..., \tau \}$, where the thresholds are satisfied with
\begin{equation}\label{Th-Winner-set}
\gamma^{1,i}_{1;1} \le \gamma^{1,j}_{1;1},\,\;\;\quad 1 \le i < j \le \tau.
\end{equation}
After the optimized pruning in the first stage, the lengths of the information bits and coded bits are decreased to $k_1 = k - 1$ and $n_1 = n - 2$, respectively.

Furthermore, the $t$-th stage optimization $2 \le t \le T$ is performed by selecting the temporary patterns ${\bf{\Pi}}^{t,i}\{1;1\}$ based on the previous pruned matrix ${\bf{H}}^{t-1,\theta(i)}_{t-1;t-1}$, where $\theta(i)$ is the parent index of $i$ in the preceding set $\hat {\bf{\Omega}}_{t-1}$. There are
\begin{equation}\label{Num-candidate-set}
|{\bf{\Omega}}_t| = \tau \times k_{t-1} \times (n_{t-1}-1)
\end{equation}
candidate pruning patterns to be evaluated by PEXIT. The patterns in set ${\bf{\Omega}}_t$ with the lowest $\tau$ thresholds are reserved as the set $\hat {\bf{\Omega}}_t=\{{\bf{\Pi}}^{t,i}\{1;1\}=\{s^{t,i}_1;p^{t,i}_1\}, i=1, 2,..., \tau \}$ for the next stage optimization.

A group of optimized pruning patterns ${\bf{\Pi}}^i\{T;T\}, i=1,..., \tau$ with length $T$ can be obtained from the set $\hat {\bf{\Omega}}_T$, after the $T$-stages optimization according to the NG ranking criterion. The $i$-th pruning pattern can be represented as follows,
\begin{equation}\label{pruning-pattern-T-rank}
\begin{aligned}
 {\bf{\Pi}}^i\{T;T\} &= \bigcup\limits_{t = 1}^T {{\bf{\Pi}}^{t,\theta^{T-t}(i)}\{1;1\}} \\
  &= \{s^{1,\theta^{T-1}(i)}_{1},...,s^{T,i}_{1}; p^{1,\theta^{T-1}(i)}_{1},...,p^{T,i}_{1}\},
\end{aligned}
\end{equation}
where the other elements before the latest $s^{T,i}_{1}$ and $p^{T,i}_{1}$ can all be traced progressively from the set $\hat {\bf{\Omega}}_{T-1}$ to $\hat {\bf{\Omega}}_1$ by the preceding indices $\theta(i),...,\theta^{T-1}(i)$.

\begin{table}[tb]
    \begin{center}
      \caption{Thresholds of the pruned base-matrices with the proposed patterns and the patterns in IEEE 802.11{\rm{n}} \label{tab:Thresholds-Pruning-Patterns-I}}
      \renewcommand{\arraystretch}{1.35}
      \renewcommand{\tabcolsep}{0.15cm}
      \begin{tabular}{|c|c|c|c|c|} \hline
%      {\bf{H}} & \multicolumn{2}{c|}{\rule[-1mm]{0mm}{4mm}11n-Z81-R1/2} & \multicolumn{2}{c|}{\rule[-1mm]{0mm}{4mm}16e-Z81-R2/3}
      {\bf{H}} & \multicolumn{2}{c|}{11n-Z81-R1/2} & \multicolumn{2}{c|}{16e-Z81-R2/3}
       \\ \hline
       Pruning & ${\bf{\Pi}}_{\bf{Opt}}\{4;4\}$ & ${\bf{\Pi}}_{\bf{11n}}\{4;4\}$ & ${\bf{\Pi}}_{\bf{Opt}}\{4;4\}$ & ${\bf{\Pi}}_{\bf{11n}}\{4;4\}$
       \\ \cline{2-5}
        patterns & \{1,2,8,10; & \{12,11,10,9; & \{4,5,8,9; & \{16,15,14,13;
       \\
        ${\bf{\Pi}}\{T;T\}$ & 5,9,19,20\} & 24,23,22,21\} & 3,20,22,23\} & 24,23,22,21\}
        \\ \hline
        $\gamma_{0;0}$(dB) & \multicolumn{2}{c|}{0.626} & \multicolumn{2}{c|}{1.472}
        \\ \hline
        $\gamma_{1;1}$(dB) & 0.571 & 0.667 & 1.523 & 1.598
        \\ \hline
        $\gamma_{2;2}$(dB) & 0.544 & 0.720 & 1.616 & 1.783
        \\ \hline
        $\gamma_{3;3}$(dB) & 0.497 & 0.780 & 1.868 & 2.039
        \\ \hline
        $\gamma_{4;4}$(dB) & 0.461 & 0.967 & 2.017 & 2.361
        \\ \hline
      \end{tabular}
    \end{center}
\end{table}
\begin{table}[tb]
    \begin{center}
      \caption{Thresholds of the pruned base-matrices with the proposed patterns and the combination patterns in  \cite{Wu2009Globecom} and \cite{Liu2009CL}} \label{tab:Thresholds-Pruning-Patterns-II}
      \renewcommand{\arraystretch}{1.35}
      \begin{tabular}{|c|c|c|c|} \hline
      \multicolumn{2}{|c|}{11n-Z81-R1/2} & \multicolumn{2}{c|}{16e-Z40-R1/2}
       \\ \hline
       ${\bf{\Pi}}_{\bf{Opt}}\{4;4\}$ & ${\bf{\Pi}}_{\bf{L\&W}}\{4;4\}$ & ${\bf{\Pi}}_{\bf{Opt}}\{0;6\}$ & ${\bf{\Pi}}_{\bf{L\&W}}\{0;6\}$
       \\ \cline{1-4}
        \{1,2,8,10; & \{3,4,6,7; & \{$\phi$; 6,14, & \{$\phi$; 13, 15
       \\
        5,9,19,20\} & 13,15,17,20\} & 16,18,20,23\} & 17,20,22,24\}
       \\ \hline
        0.461(dB) & 0.922(dB) & 1.551(dB) & 1.573(dB)
       \\ \hline
      \end{tabular}
    \end{center}
\end{table}

The threshold of any sub-pattern ${\bf{\Pi}}^i\{\alpha; \beta\}, 0 \le \alpha \neq \beta \le T$ of pattern ${\bf{\Pi}}^i\{T;T\}$ is usually not the lowest one. However, a good performance is still provided for the arbitrary sub-patterns of the pruning pattern ${\bf{\Pi}}^i\{T;T\}$, which is shown in the simulation results. Without loss of generality, we use the pruning pattern ${\bf{\Pi}}^1\{T;T\}$ as the final pruning pattern ${\bf{\Pi}}\{T;T\}$ for the base-matrix ${\bf{H}}$.

\begin{figure}[!t]
    \begin{center}
      \epsfxsize=7.0in\includegraphics[scale=0.482]{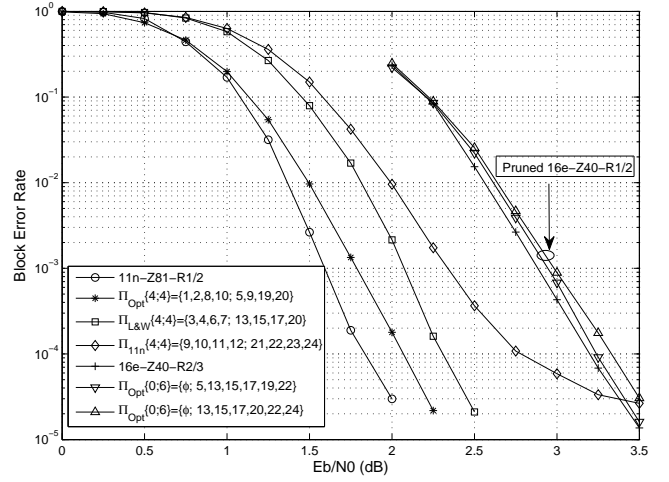}
      \caption{Error performance of 11n-Z81-R1/2 and 16e-Z40-R1/2 codes pruned by different patterns.
      \label{fig:Sim-R12-1944}}
    \end{center}
\end{figure}

\begin{figure}[!h]
    \begin{center}
      \epsfxsize=7.0in\includegraphics[scale=0.482]{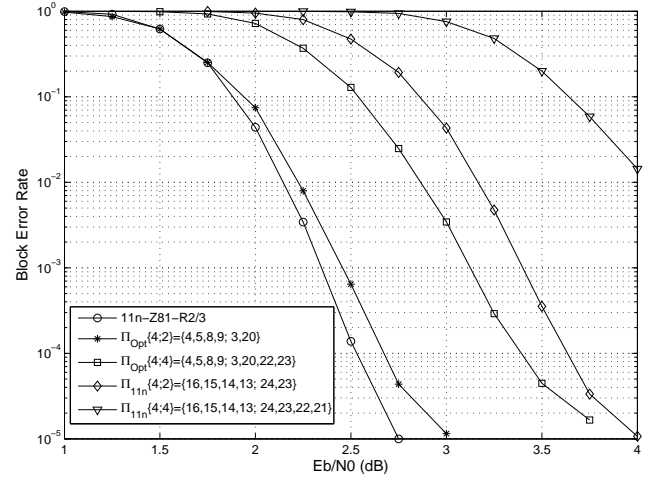}
      \caption{Error performance of 11n-Z81-R2/3 code pruned by different patterns. \label{fig:Sim-R23-1944}}
    \end{center}
\end{figure}

\section{Simulation Results}
The structured LDPC codes specified in IEEE 802.11n and 802.16e systems are used for the performance comparison with different pruning patterns. For example, the two rate-1/2 LDPC codes with lifting factor $Z=81$ and $Z=40$ in IEEE 802.11n and 802.16e systems are denoted by 11n-Z81-R2/3 and 16e-Z40-R1/2, respectively. Since there are no joint puncturing and shortening optimized pruning patterns for the structured LDPC codes in other existing methods, we only compare the simulation results between the pruning patterns optimized by our proposed method (${\bf{\Pi}}_{\bf{Opt}}$), the combination pruning patterns in \cite{Liu2009CL,Wu2009Globecom} (${\bf{\Pi}}_{\bf{L\&W}}$), and those defined in \cite{IEEE_P802_11n} (${\bf{\Pi}}_{\bf{11n}}$). In all simulations, the belief propagation (BP) algorithm is used and the maximum number of iterations is set to 100 for LDPC decoding.

The threshold comparisons of different pruned LDPC codes, 11n-Z81-R1/2, 11n-Z81-R2/3 and 16e-Z40-R1/2, are given in Table \ref{tab:Thresholds-Pruning-Patterns-I} and Table \ref{tab:Thresholds-Pruning-Patterns-II}, where the pruning patterns ${\bf{\Pi}}_{\bf{11n}}\{4;4\}$ and ${\bf{\Pi}}_{\bf{Opt}}\{4;4\}$ are proposed in IEEE 802.11n standard and optimized by NG ranking criterion with $T=4$, respectively. The thresholds $\gamma_{\alpha;\beta}, \alpha,\beta \le 4$ are evaluated by the pruned base-matrices with sub-patterns ${\bf{\Pi}}_{\bf{Opt}}\{\alpha;\beta\}$ and ${\bf{\pi}}_{\bf{11n}}\{\alpha;\beta\}$, which consist of the first $\alpha$ shortened and $\beta$ punctured components in patterns ${\bf{\Pi}}_{\bf{Opt}}\{4;4\}$ and ${\bf{\Pi}}_{\bf{11n}}\{4;4\}$. The final code rate pruned by the pattern ${\bf{\Pi}}_{\bf{Opt}}\{\alpha,\beta\}$ or ${\bf{\Pi}}_{\bf{11n}}\{\alpha,\beta\}$ is $(k-\alpha)/(n-\alpha-\beta)$. For example, the highest code rates of the two LDPC codes in the table \ref{tab:Thresholds-Pruning-Patterns-I} can achieve $0.6$ and $0.8$, respectively, after pruned with the pattern parameters $\alpha=0$ and $\beta=4$.

From the numeric results presented in Table \ref{tab:Thresholds-Pruning-Patterns-II}, we can see that there is a noticeable gap between the threshold of the pattern ${\bf{\Pi}}_{\bf{Opt}}\{4;4\}$ and that of the direct combination of patterns ${\bf{\Pi}}_{\bf{L\&W}}\{4;4\}$, which are individually optimized for shortening \cite{Liu2009CL} and puncturing \cite{Wu2009Globecom}, respectively. Since the punctured nodes in our pattern are selected from both the systematic part and the parity part of the codeword, the threshold of our progressively puncturing pattern ${\bf{\Pi}}_{\bf{Opt}}\{0;6\}$ for code 16e-Z40-R1/2 is slightly better than that of the pattern ${\bf{\Pi}}_{\bf{L\&W}}\{0;6\}$ in \cite{Wu2009Globecom}, when the number of the punctured nodes is limited. If the puncturing process for rate higher than 2/3 is further carried out, the puncturing pattern optimized in \cite{Wu2009Globecom} will be more efficient than ours. Fortunately, the LDPC codes with different code rates, such as 1/2, 2/3, 3/4 and 5/6, are all specified in IEEE 802.16e, so the number of the punctured nodes usually is limited.

Fig. \ref{fig:Sim-R12-1944} shows the error performance of the 11n-Z81-R1/2 code pruned by the different pruning patterns. There is only small performance loss for the pruned code with the pattern ${\bf{\Pi}}_{\bf{Opt}}\{4;4\}$, although the real transmission length decreases from 1944 bits to 1296 bits. Moreover, noticeable performance gain can be achieved with our optimized pattern compared with the patterns ${\bf{\Pi}}_{\bf{L\&W}}\{4;4\}$ and ${\bf{\Pi}}_{\bf{11n}}\{4;4\}$. The 16e-Z40-R1/2 code used in \cite{Wu2009Globecom} with only puncturing pattern ${\bf{\Pi}}_{\bf{Opt}}\{0;6\}$ also slightly outperforms that with the pattern ${\bf{\Pi}}_{\bf{L\&W}}\{0;6\}$ in \cite{Wu2009Globecom}, where the transmission rate is same as that of the 16e-Z40-R2/3 code in \cite{IEEE_P802_16e}.

The performance comparisons of the high-rate code, 11n-Z81-R2/3, with different pruning patterns are demonstrated in Fig. \ref{fig:Sim-R23-1944}. The error performance of the pruned code with our optimized patterns, ${\bf{\Pi}}_{\bf{Opt}}\{4;4\}$ and ${\bf{\Pi}}_{\bf{Opt}}\{4;2\}$, is always better than that with the patterns ${\bf{\Pi}}_{\bf{11n}}\{4;4\}$ and ${\bf{\Pi}}_{\bf{11n}}\{4;2\}$. The transmission rate can maintain the original code rate 2/3, when the sub-patterns with $\alpha=4$ and $\beta=2$ are selected. It can also be seen from Fig. \ref{fig:Sim-R23-1944} that the performance of the pruned codes with pattern ${\bf{\Pi}}_{\bf{Opt}}\{4;2\}$ are very close to that of the unpruned code, although the transmission lengths are much shorter than the original 1944 bits.

\section{Conclusion}
We propose an efficient optimization scheme for the pruning of structured LDPC codes, which can be evaluated by the PEXIT analysis according to the protographs mapped from the base-matrices. A $T$-stage progressive optimized pruning pattern can be obtained according to the NG ranking criterion, where any sub-pattern ${\bf{\Pi}}_{\bf{Opt}}\{\alpha;\beta\}, \alpha \le T, \beta \le T $ is composed of the first $\alpha$ shortened and $\beta$ punctured components in the pattern ${\bf{\Pi}}_{\bf{Opt}}\{T;T\}$. In terms of the structured LDPC codes in IEEE 802.11n, both the numerical analysis and simulation results show that our optimized pruning patterns apparently outperform the existing pruning patterns in IEEE 802.11n standard.

%\bibliographystyle{IEEEtran}
%\bibliography{protograph}

%\clearpage
%\newpage
\end{document}